\documentclass[pra,aps,notitlepage,nofootinbib,secnumarabic,10pt]{revtex4-2}

\usepackage[bookmarks=false,colorlinks=true]{hyperref}
\usepackage{amsmath,amsfonts,amssymb}
\usepackage{physics}
\usepackage{setspace}
\usepackage{mathtools}

\newcommand{\be}{\begin{equation}}
\newcommand{\ee}{\end{equation}}
\newcommand{\sfrac}[2]{{\textstyle\frac{#1}{#2}}}
\newcommand{\sbinom}[2]{{\textstyle\binom{#1}{#2}}}

\begin{document}

\title{Dunkl symplectic algebra in generalized Calogero models}
\author{Tigran Hakobyan}
\email{tigran.hakobyan@ysu.am, hakob@yerphi.am}
\affiliation{Yerevan State University, 1 Alex Manoogian Street, Yerevan, 0025, Armenia}
\affiliation{Alikhanyan Natianal Laboratory, 2 Aliknanyan br. Street, Yerevan, 0036, Armenia}
\begin{abstract}
We study the properties of the symplectic $sp(2N)$ algebra deformed using Dunkl
operators, which describe the dynamical symmetry  of the generalized $N$-particle quantum Calogero model.
It contains a symmetry subalgebra formed by the deformed unitary generators
as well as the (nondeformed) $sl(2,R)$ conformal  subalgebra.
An explicit relation among
the deformed symplectic generators is derived.
Based on the matching between the  Casimir elements of the conformal spin
and  Dunkl angular momentum algebras,
the independent wavefunctions of the both the standard and generalized Calogero models, expressed in terms
of the deformed spherical harmonics, are classified according to infinite-dimensional lowest-state $sl(2,R)$ multiplets.
Meanwhile, any polynomial integral of motion of the (generalized) Calogero-Moser
model generates a finite-dimensional highest-state conformal  multiplet  with descendants
expressed via the Weyl-ordered product in quantum field theory.
\end{abstract}

\maketitle


\section{Introduction}

Most many-body systems with particle interactions do not possess exact solutions in either classical or quantum mechanics.
They are typically amenable to numerical simulations and analytical approximations.
However, in one dimension, there exists a remarkable  family of integrable  $N$-particle
systems based on the Calogero model. The latter
consists of identical particles moving in one space dimension with pairwise inverse-square interaction.
The quantum  Calogero model is described by the following Hamiltonian which possesses an
exact solution  \cite{cal}:
\begin{equation}
\label{H}
H=\frac12\sum_{i=1}^N(p_i^2+x_i^2) + \sum_{i<j}\frac{g(g\mp 1)}{(x_i-x_j)^2}.
\end{equation}
The presence or absence of the confining harmonic potential (as well as any other potential depending solely on the radial coordinate)
does not affect the integrability.
The particle mass, frequency, and Planck's constant are set to unity in the above Hamiltonian.
The $-/+$ sign in the potential coupling constant
is chosen for bosons/fermions, respectively. The corresponding term is absent in the classical
case.
Many properties of the classical and quantum systems are similar. In particular,
in the absence of the harmonic potential, the Liouville  integrals of
the classical Hamiltonian were constructed by  the Lax matrix method \cite{moser}.
Later, a similar method  was applied in order to establish the integrability
the related quantum model \cite{hikami}.
In the current article, the quantum model is considered, with $p_j = -\imath \partial/\partial x_j$.
More general inverse-square potentials based on finite reflection groups, trigonometric functions
\cite{suther}, spin exchanges, etc.,
are also integrable. 
(See the reviews~\cite{rev-olsh,rev-poly,rev-harut,rev-edin} and references therein.)

The family of conservation laws in such systems is much more diverse and  rich.
In particular, aside from the $N$ commuting integrals, the Calogero systems with rational potentials
possess $N-1$
 additional constants of motion, making them
 maximally superintegrable  both in the classical \cite{woj83} and quantum \cite{kuznetsov,gonera98}
 cases.
This property
results in a significant degeneracy of the energy levels of quantum system and drastically simplifies  the solution.
The (super)integrability is robust under certain external potentials and
persists for angular parts of the related Hamiltonians, as well as for systems defined on surfaces with constant curvature
\cite{CalCoul,runge,flp,fh}.

Another closely related and intriguing property of  the described models
is  the elimination of the inverse-square potential through a nonlocal unitary transformation,
rendering the many-body systems equivalent to their noninteracting counterparts
 \cite{gur99,gonera99}.
In the quantum case, the equivalence becomes even more transparent when
the particle interaction term is incorporated into a nonlocal covariant derivative with exchange operators
 introduced by Dunkl \cite{dunkl}.
A generalized interaction, obtained in this way, contains the particle exchanges and  is reduced
to the original Calogero potential
 for the identical particles
\cite{poly92,brink}.
 Furthermore, recent discoveries have shown that quantum systems with even more general
 particle exchange interactions. They also can be solved in the thermodynamic limit and possess
 scattering matrices obeying  the Yang-Baxter equation. However, this family
 comprise (probably) nonintegrable systems with  backscattering   \cite{poly20}.

 The described correspondence between the noninteracting and interacting models extends further to their symmetries.
 In particular, the Dunkl deformations of the momentum and angular momentum
 serve as invariants for the unbounded Calogero Hamiltonian with particle exchanges,
 providing an argument for superintegrability at the level of the Dunkl operator. The Dunkl angular momentum algebra
 also encompasses the complete symmetries of the generalized angular Calogero model
\cite{fh}.
The same deformation applied to the symmetry generators of the
$N$-dimensional isotropic oscillator
results in the Dunkl-deformed version of the related unitary  Lie algebra $u(N)$, which
 describes the symmetry of the generalized Calogero model \cite{fh}.
In case of  the confining Coulomb potential,
the Dunkl deformation of the $so(N +1)$ algebra is obtained \cite{fh2}.
The latter encompasses all symmetries of the generalized Calogero-Coulomb model and includes the Dunkl analogs of the Runge-Lenz vector and angular momentum tensor
 \cite{runge}.

The Dunkl deformation procedure largely preserves or only slightly alters many crucial properties
of the original symmetries.
Notably, the commutation and algebraic relations among the generators,
the Casimir elements, and their action on the wavefunctions closely resemble
their non-deformed counterparts. However, it is worth noting that the deformed generators do not constitute a Lie algebra based solely on commutation but rather form a
quadratic algebra in conjunction with the
exchange operators.

The described mapping extends further
to the spectrum generating algebras, also known as dynamical symmetries.
For the $N$-dimensional isotropic oscillator, the dynamical symmetry
can be described by the semidirect product of Weyl%
\footnote{
The continuous  Weyl group $W_N$ and the corresponding algebra $w_N$
is generated by the creation-annihilation operators.
The latter is similar to the Heisenberg algebra with imaginary unit replaced by the unit.
It differs from the discrete Weyl group related with the finite reflection groups, which
in the current article is the permutation group $S_N$.
}
 and symplectic groups, denoted as $W_N \rtimes Sp(2N)$.
These groups are generated by linear and bilinear combinations of the creation-annihilation operators \cite{wybourne}.
In this article, we study the Dunkl deformation of the symplectic Lie algebra, with particular attention to explicit commutation
relations among its generators.
The Dunkl $sp(2N)$  algebra consists of the deformed unitary subalgebra, which encompasses the symmetries
of the generalized Calogero Hamiltonian, as well as the spectrum-generating part
that maps between different energy levels. Additionally, it contains the standard $sl(2,R)$ conformal subalgebra.
Notably, the quadratic invariants of both the conformal spin and Dunkl angular momentum coincide. As a result,
the eigenfunctions of the generalized Calogero Hamiltonian in spherical coordinates, expressed in terms of the deformed spherical harmonics,
unify in the lowest-weight $sl(2,R)$ representations.
The second quantum number in the wavefunction varies within an individual multiplet and describes the projection of the conformal spin, while the remaining parameters
characterize the representation.
For the (generalized) Calogero-Moser model, any integral of motion generates a finite-dimensional highest weight $sl(2,R)$ multiplet, with the descendants expressed in
terms of the Weyl-ordered product in quantum
field theory.
The highest states  in a product of two or more such multiplets
produce additional integrals of motion, which are responsible for
the superintegrability.

The article   is organized as follows.
In Sect.~\ref{sec:model},   the extended version of the Calogero model, which
is based on the Dunkl operators, is briefly reviewed. In Sect.~\ref{sec:symplectic}, we construct and study
the Dunkl analog of the symplectic group generators.
Together with the Dunkl creation-annihilation operators,
they form a Dunkl deformation of the $w_N\rtimes sp(2N)$ algebra.
In Sect.~\ref{sec:wavefunctions}, we examine the  behaviour of the eigenfunctions
of the generalized Calogero system under the conformal group, utilizing
the Dunkl spherical harmonics.  Sect.~\ref{sec:integrals}
reveals the conformal algebra structure of the commuting integrals of motion
of the Calogero-Moser system. Using the momentum sum rules and Weyl ordering, the
additional integrals are constructed.
The results are summarized in Conclusion. 


%

\section{Generalized Calogero model}
\label{sec:model}
%

\subsection{Hamiltonian and Dunkl operators}
The  solution and integrals of motion of the quantum Calogero system \eqref{H}
may be formulated in an elegant way after slight but nontrivial modification.
The expanded version  involves the particle exchanges
in the potential \cite{poly92,brink}:
\begin{equation}
\label{Hgen}
H=\frac12\sum_{i=1}^N(p_i^2+x_i^2) + \sum_{i<j}\frac{g(g-s_{ij})}{(x_i-x_j)^2}.
\end{equation}
Here $g>0$ is an attractive coupling constant, and the operator $s_{ij}$ permutes the $i$-th and $j$-th
 particles while leaving the others untouched. For  identical bosons or fermions, the
 eigenfunctions are symmetric or antisymmetric, respectively. Therefore, $s_{ij}=\pm1$, and above
 Hamiltonian is reduced to the original Calogero model \eqref{H}.

Moreover, the inverse-square potential  may be absorbed by
a covariant momentum allowing us to rewrite the above Hamiltonian
as follows:
\begin{equation}
\label{pi}
H=\frac12\sum_{i=1}^N(\pi_i^2+x_i^2),
\qquad
\pi_i=-\imath \nabla_i.
\end{equation}
The covariant momentum $\pi_i$ is nonlocal and depends on particle
permutations.
It is defined by the Dunkl operator \cite{dunkl}:
\begin{equation}
\label{du}
\nabla_i=\partial_i-\sum_{j\ne i} \frac{g}{x_i-x_j}s_{ij}.
\end{equation}
The system \eqref{Hgen}, \eqref{pi} is known as the generalized Calogero model,
often referred to as  a Dunkl oscillator.

The components of the Dunkl momentum  mutually commute, resulting in a
 flat connection. However, the commutations with
coordinates are more  intricate compared to the standard canonical commutation
relations:
\begin{equation}
\label{comm}
[\nabla_i,\nabla_j]=0,
\qquad
[\nabla_i,x_j]=S_{ij}.
\end{equation}
The last equation contains a matrix formed by the particle-exchange
operators,
\begin{equation}
\label{Sij}
S_{ij}=\begin{cases}
-g s_{ij} & i\ne j,
\\
1+g\sum_{k\ne i} s_{ik} & i=j.
\end{cases}
\end{equation}
 In the free-particle limit ($g\to0$), it   reduces  to   the Kronecker delta: $S_{ij}\to \delta_{ij}$.
 Then  the coordinates, momentum and the unit (representing here the Planck's constant) form
the  Heisenberg Lie algebra.
Therefore, the algebra \eqref{comm}, \eqref{Sij} is often referred to in the literature as an
$S_N$-extended Heisenberg  algebra \cite{brink'}.
It can be considered also as a deformation
of the formal power series in coordinates and momenta (which constitute the universal enveloping algebra
of the  Heisenberg Lie algebra)
to an associative algebra formed by the  elements $P Q s$,
where  $P=P(x_1,\dots x_N)$ and $Q=Q(\nabla_1,\dots, \nabla_N)$ are polynomials
 and $s$ belongs to the symmetric group $S_N$. Note that the elements $x_i$, $\nabla_i$, and
 $S_{ij}$ do not form a Lie algebra for a nontrivial coupling $g$.
An abstract algebraic structure which describes  the mentioned deformation is called
the rational Cherednik algebra associated with the symmetric group \cite{cherednik,etingof},
and the Dunkl operators with coordinates generate its  particular representation.

It is worth noting that there is a trigonometric version of the Dunkl operators
\cite{hechman91,cherednik}. The related dynamical system is the Calogero-Sutherland model \cite{suther}.
Its version  with particle exchanges also has the simple expression \eqref{pi}.
The exchange operator formalism was also applied to integrable spin chains with long-range interactions
\cite{poly93, minahan93}. 
Moreover, an appropriate combination of spin and particle exchanges
gives rise to integrable models with spin and dynamical degrees of freedom \cite{poly92},
which reduce to lattice systems after freezing the coordinates at equilibrium \cite{poly93}.
These composite systems exhibit essential features of integrable models:
Yangian symmetry, Lax pair equation, and monodromy matrix obeying the Yang-Baxter equation \cite{bernard}.
Despite the variety, all these algebraic properties are based on the Dunkl-operator
framework \cite{lam-serb}.

As was already mentioned in Introduction, the integrability of the Calogero model \eqref{H}
can be also established  by the quantum inverse scattering approach \cite{hikami}.
For the quantum systems, in order to derive the conserved quantities from
the Lax-matrix power, one usually applies the summation over all matrix elements instead
of taking the trace. When using the Dunkl-operator method, one must restrict to the
indistinguishable-particle sector which eliminates the permutation operators \cite{poly92}.
Although both approaches are based on different algebraic structures, there is a correspondence
between them \cite{chalykh}.
In particular, the commuting integrals of motion obtained by  both
methods turn out to be equivalent \cite{wadati2}.

\subsection{Creation-annihilation operators}
The generalized Calogero Hamiltonian \eqref{pi} bears resemblance to
the isotropic oscillator, making it amenable to straightforward
and powerful tools for studying the spectrum, integrals of motion, and more.
Notably, the Hamiltonian can be expressed in terms of Dunkl
analogs of the lowering and raising operators for energy levels
\cite{poly92,brink}:
\begin{equation}
\label{Ha}
H=\sum_{i=1}^Na_i^+a_i+\frac12 N-S,
\end{equation}
\begin{equation}
\label{apm}
a_i^\pm=\frac1{\sqrt2}(x_i\mp\nabla_i).
\end{equation}
They serve the same role as the particle
creation and annihilation operators in the field-theoretical treatment.
The final component in the Hamiltonian \eqref{Ha}, which disappears
in the free-particle limit,
is  the permutation-invariant
sum over all pairwise exchanges \cite{fh}:
\begin{equation}
\label{S}
S=\sum_{i<j}S_{ij},
\qquad
[S,S_{ij}]=0.
\end{equation}

The ladder operators obey similar relations as the coordinate and
Dunkl derivative:
\begin{equation}
\label{com-apm}
[a_i,a_j^+]=S_{ij},
\qquad
[a_j,a_j]=[a_j^+,a_j^+]=0.
\end{equation}
They constitute a Dunkl version of the Weyl algebra
$w_N$. The latter corresponds to the $g=0$ case of above relations
when $S_{ij}=\delta_{ij}$. The above relations are completed
by commutations with permutation operators:
\begin{equation}
\label{com-apmS}
S_{ij}a^\pm_j=a^\pm_jS_{ij},
\qquad
[S_{ij},a^\pm_k]=0
\end{equation}
with distinct values of all three indices.

The standard spectrum-generating commutation rules
apply to the Calogero Hamiltonian with particle exchanges:
\begin{equation}
\label{comHa}
[H,a_i^\pm]=\pm a_i^\pm.
\end{equation}

The total Dunkl momentum, however, does not involve particle exchanges and
therefore reduces to the usual momentum. Consequently, the
center-of-mass creation-annihilation operators obey the standard
one-particle Weyl algebra commutation relation:
\begin{equation}
\label{bpm}
A^\pm=\frac1{\sqrt{N}}\sum_i a_i^\pm,
\qquad
[A,A^+]=1.
\end{equation}

\subsection{Energy spectrum and eigenstates}

Similar to the isotropic oscillator, the energy levels and eigenstates of the generalized Calogero model
are constructed algebraically.

The lowering operators annihilate the ground state, which
 is given by the following wavefunction (with the normalization constant omitted)
 \cite{brink}:
\begin{equation}
\label{psi0}
|0\rangle=e^{-\frac12r^2}\prod_{i<j}|x_i-x_j|^g,
\qquad
a_i|0\rangle=0,
\end{equation}
where the radial coordinate defined is by the equation  $r^2=\sum_i x_i^2$.
The lowest energy is given by the following expression:
\begin{equation}
\label{E0}
E_0=\frac12 g N(N-1)+\frac12N.
\end{equation}
Besides the standard oscillatory energy, which contributes one-half per each degree of freedom,
it also includes the interaction part.
The interaction assigns the value $g$ to each particle pair. According to Eq.~\eqref{comHa},
excitations above the lowest state are generated by monomials in raising operators,
resulting in an equidistant and highly degenerate spectrum, similar to the oscillator case:
\begin{equation}
|{\bf n}\rangle
=
(a_1^+)^{n_1}\dots (a_N^+)^{n_N}|0\rangle,
\qquad
\label{En}
E_{\bf n}=E_0+\sum_{l=1}^N n_l.
\end{equation}
It is worth noting that the particle interaction merely
shifts all levels by the same value given by the $g$-term in the ground-state energy
\eqref{E0}.

In general, these states describe distinguishable particles since they vary under particle exchanges.
To obtain identical bosons (fermions) from them, an additional (anti)symmetrization procedure is required.
This reduction results in the original Hamiltonian  \eqref{H}.
Alternatively, one can apply (anti)symmetric polynomials in raising operators to the ground state.
For instance, using power sums,
\begin{equation}
\label{Al}
 A_l^\pm=\sum_{i=1}^N (a^\pm_i)^l,
 \end{equation}
the related (unnormalized) bosonic eigenstates are labeled by $N$ non-negative integers,
${\bf k}=\{k_1,\dots,k_N\}$,
and given by the formula
\cite{cal,flp}
\begin{align}
\label{ksym}
|{\bf k}\rangle_\text{sym}
=
(A_1^+)^{k_1}(A_2^+)^{k_2}\dots (A_N^+)^{k_N}|0\rangle.
\end{align}
The  corresponding energy levels are highly degenerate and given by
the following expression:
\begin{equation}
\label{Ekb}
E_{\bf k}=E_0+m_{\bf k}
\qquad
\text{with}
\qquad
m_{\bf k}=\sum_{l=1}^N lk_l
\end{equation}
being the degree of the symmetric polynomial in
$a_i^+$ operators. Note that the above eigenfunctions
may be obtained also using the  quantum  Lax operator algebra  \cite{wadati}.

The first quantum number $k_1$ describes the center-of-mass
energy of the system. The related symmetrized operator \eqref{Al}
is proportional to the  Weyl algebra generator \eqref{bpm}:
$A_1^\pm=\sqrt{N}A^\pm$.
The
annihilation operator
reduces the energy level by one, and its action
on the energy eigenstates can be easily calculated:
\begin{equation}
\label{A1}
A_1^-|{\bf k}\rangle_\text{sym}
=
\sum_{l=1}^{N}
lk_l|{\bf k}-{\bf e}_l+{\bf e}_{l-1}\rangle_\text{sym}.
\end{equation}
Here ${\bf e}_l$ is the  $l$-th standard unit vector in $N$ dimensional
space with the components $\delta_l^i$ for $1\le l\le N$, and
${\bf e}_0=0$.
The states  with a negative coefficient $k_i$
must be  eliminated from  the above sum.

%

%
%
%
%

\section{Dunkl deformed symplectic algebra}
\label{sec:symplectic}

\subsection{Dunkl deformation of $sp(2N)$ generators}
One can enhance the ladder operators \eqref{apm} by including their bilinear combinations,
\begin{equation}
\label{EF}
E_{ij}=a^+_ia_j,
\qquad
F_{ij}=a_ia_j,
\qquad
F_{ij}^+=a_i^+a_j^+.
\end{equation}
Clearly, among them, only $N(2N+1)$ are distinct. These elements generate a Dunkl deformation
of the symplectic Lie algebra $sp(2N)$. When combined with the operators $a^\pm_i$, they
collectively form
a Dunkl deformation of the dynamical symmetry algebra for the generalized Calogero model \eqref{Hgen}.
The latter extends the   dynamical symmetry group $W_N\rtimes Sp(2N)$ associated with the isotropic
harmonic oscillator \cite{wybourne}.

When considering the original Calogero model \eqref{H} instead of the generalized one
\eqref{Hgen}, we typically work with symmetric polynomials in the elements $E_{ij}$ and $F^\pm_{ij}$.
This choice is made because the original system often involves identical particles, such as bosons or fermions.

 Looking ahead, the spectrum-generating algebra for the original system consists of independent polynomials taken from the set
$\sum_{i=1}^N(a_i^+)^ka_i^l$,
 which includes the Hamiltonian itself and the conformal generators discussed below.
These polynomials have rather intricate commutation relations among them
 \cite{jonke01}.

\subsection{Deformed unitary symmetry subalgebra}
\label{sec:symmetry}
The first set of generators in \eqref{EF} is formed by the $N^2$ elements $E_{ij}$
and constitutes the symmetry algebra of the
generalized Calogero Hamiltonian \cite{turbiner,fh}. Each one transposes its indices
under Hermitian conjugate,
$$
[H,E_{ij}]=0, \qquad
E_{ij}^+=E_{ji}.
$$
They constitute the Dunkl-operator deformation of the standard $gl(N)$  basis
 and obey the following commutation rules \cite{fh}:
\begin{equation}
\label{comE}
[E_{ij},E_{kl}]=E_{il}S_{jk}-S_{il}E_{kj}+[S_{kl},E_{ij}].
\end{equation}
%
The explicit commutation relations with the Dunkl ladder operators are:
\begin{equation}
\label{comEa}
[E_{ij},a_k^+]=a_i^+S_{jk},
\qquad
[E_{ij},a_k]=-a_jS_{ik}.
\end{equation}

The symmetry generators split into the antisymmetric and symmetric parts,
giving rise to the Dunkl angular momentum and Fradkin tensors, respectively  \cite{kuznetsov,feigin,fh,francisco}.
Both of them are Hermitian and represent the Dunkl analog of the $U(N)$ generators:
\begin{equation}
\label{Lij}
L_{ij}=-\imath (E_{ij}-E_{ji}),
\qquad
I_{ij}= E_{ij}+E_{ji}.
\end{equation}
In the absence of the interaction part ($g=0$), they generate the unitary
symmetry of the isotropic oscillator.

The Dunkl angular momentum
components  are closed under the commutation with coefficients that depend
on the particle exchanges \cite{fh}:
$$
[L_{ij},L_{kl}]=-\imath(L_{il}S_{jk}+L_{jk}S_{il}-L_{ik}S_{jl}+L_{jl}S_{ik}).
$$
The Casimir element, which commutes with above generators
and modifies  the usual angular momentum squared,
 yields the angular part of the Calogero model (see also Eq.~\eqref{H0} below) \cite{fh},
\begin{equation}
\label{Homega}
\begin{aligned}
H_\Omega=L^2+S(S-N+2),
\qquad
L^2=\sum_{i<j}L_{ij}^2,
\qquad
[H_\Omega,L_{ij}]=0.
\end{aligned}
\end{equation}
Using the commutation relations among the coordinates and Dunkl
operators, the Dunkl angular momentum squared can be expressed as follows \cite{fh}:
\begin{equation}
\label{Homega'}
H_\Omega=-r^2\nabla^2+r\partial_r(r\partial_r+N-2)
\end{equation}
where $r=\sqrt{x^2}$ is the radial coordinate. Here, $x^2=\sum_i x_i^2$, and $\nabla^2=\sum_i\nabla_i^2$.
It is worth noting that the Dunkl angular momentum has also
emerged as the symmetry of the Dunkl-operator extended version of the free relativistic particle  \cite{Bie}.

The Dunkl $u(N)$ algebra also possesses a Casimir element,
which, similar to the isotropic oscillator, is expressed in terms of the Hamiltonian:
$$
C=\sum_{i,j}E_{ij}(E_{ji}+S_{ji})=H^2-(S-\sfrac12 N)^2.
$$
The formula above is a result of a quadratic relation, known as the crossing relation, among the
generators of the Dunkl unitary algebra \cite{fh}:
\begin{equation}
\label{EE-ES}
E_{ij}(E_{kl}+S_{kl})=E_{il}(E_{kj}+S_{kj}).
\end{equation}

It is instructive to uncover the behavior of the described symmetry generators \eqref{Lij}
in  the weak harmonic potential limit. Throughout the current article, the corresponding
frequency value is set to unity  without loss of generality: $w=1$.
 This parameter can be recovered
by redefining the ladder operators \eqref{apm} and the Hamiltonian \eqref{Ha}
as follows:
\begin{equation}
\label{Hw}
a_i\to \sqrt{\frac{w}{2}} x_i\mp \frac{1}{\sqrt{2w}}\nabla_i,
\qquad
H\to w H =\frac12 \nabla^2 +\frac{w^2}{2}x^2.
\end{equation}
It is also appropriate to renormalise  the Dunkl analog of
the Fradkin tensor
via  $I_{ij}\to w I_{ij}$.   Then, the $w=0$ part of the conservation condition $[H,I_{ij}]=0$
would give the evident equation  $[\nabla^2,\nabla_i\nabla_j]=0$.
The Dunkl angular momentum does not depend on
$w$, and its conservation would provide the relation $[\nabla^2,L_{ij}]=0$.
The last two equations support the fact that both the Dunkl momentum $\pi_i=-\imath\nabla_i$ and
Dunkl angular momentum generate an entire symmetry algebra of the generalized
Calogero system without confining harmonic potential (generalized
Calogero-Moser model, see Eq.~\eqref{H0} below). The explicit algebraic
relations among them are known \cite{fh}. In $g=0$
limit, they generate the Euclidean group $E(N)\equiv ISO(N)$
which describes the symmetries of the free-particle system. The latter
significantly differs from the unitary symmetry $U(N)$ of the
isotropic oscillator.

\subsection{Spectrum generating part}
Finally, the remaining $N(N+1)$ operators from the definition \eqref{EF}
form Dunkl analogs of the ladder generators within the symplectic group $Sp(2N)$.
The corresponding matrices are symmetric and
 map between the next-to-nearest neighboring  energy levels of the
 generalized Calogero Hamiltonian \eqref{Hgen}:
\begin{equation}
\label{comHF}
F^\pm_{ij}=F^\pm_{ji},
\qquad
[H,F^\pm_{ij}]=\pm2 F^\pm_{ij}.
\end{equation}

The commutation relations  with the Dunkl creation-annihilation
operators are analogous to those for the symmetry generators \eqref{comEa}.
The nontrivial commutators among them
can be readily obtained using the algebra \eqref{com-apm},
\begin{equation}
\label{comFa}
\begin{aligned}
[F_{ij},a_k^+]&=S_{ik}a_j+a_iS_{jk},
\qquad
[F^+_{ij},a_k]&=-S_{ik}a^+_j-a^+_iS_{jk}.
\end{aligned}
\end{equation}

The remaining commutators among the $F$ operators, as well as
 between the $F$ and $E$ operators, are more complex.
For \emph{distinct} values of the four indices, Wick's theorem holds for a string
composed from the product of four ladder operators with the contraction among the pairs
$a_i$ and $a_k^+$ provided by the exchange operator $S_{ik}$. The latter
behaves like a number since it commutes with the other operators  in that string.
For example, we have:
$$
a_ia_ja^+_ka_l=a_k^+a_la_ia_j + a_ia_l S_{jk}+a_ja_l S_{ik}.
$$
As a result, the only nontrivial commutators are:
\begin{align*}
&[F_{ij},F_{kl}^+]=E_{ki}S_{jl}+E_{kj}S_{li}+E_{li}S_{kj}+E_{lj}S_{ik},
\\
&[F_{ij},E_{kl}]=F_{jl}S_{ik} +F_{il}S_{jk},
\end{align*}
and their conjugates. 

For \emph{general} values of indices, the commutation relations among the generators
may be expressed as:
\begin{align}
 [F_{ij},F_{kl}^+] =&\, S_{ik}(E_{lj}+S_{lj}) + E_{ki}S_{lj} +a_iS_{jk}a_l^+
 +a_k^+ S_{il} a_j,
 \label{comFF}
\\
 [ F_{ij},E_{kl}]  = &\, S_{ik} F_{jl} +a_iS_{jk}a_l
  \label{comFE}
\end{align}
with the last equation supplemented by its Hermitian conjugate.
The commutator with the Dunkl angular momentum \eqref{Lij} is inherited from
Eq.~\eqref{comFE} and its conjugate:
\begin{equation}
\label{comFL}
[F_{ij}^\pm,L_{kl}]=-\imath(S_{ik}F_{jl}^\pm+F_{il}^\pm S_{jk}-S_{il}F^\pm_{jk}-F_{ik}^\pm S_{jl}).
\end{equation}

In formulas \eqref{comFF}, \eqref{comFE}, the terms that  involve bilinear combinations of Dunkl creation and annihilation operators can always be re-expressed
in terms of the $E$ or $F$ generators. This can be achieved by moving the permutation operators $S_{ij}$ to either  the
left or the right.
In particular, when the four indices are split into two equal pairs with values $i\ne j$, the relation
\eqref{comFF}  simplifies as follows:
\begin{align}
\label{comFiiFkk}
[F_{ii},F_{kk}^+]=&(E_{ki}+E_{ik}+E_{kk}+E_{ii}+S_{ii})S_{ik}+g^2,
\\
\label{comFik}
[F_{ik},F_{ik}^+]=&(2E_{ii}+S_{ii})S_{ik}+S_{ii}(E_{kk}+S_{kk})
  +E_{ii}S_{kk}.
\end{align}
 Then the next expression  \eqref{comFE} turns into the following
couple:
\begin{align}
\label{comFiiEkk}
[F_{ii},E_{kk}]&=(F_{ik}+F_{ii})S_{ik},
\\
\label{comFikEik}
[F_{ik},E_{ik}]&=F_{ii}S_{ik}   + S_{ii}F_{kk}.
\end{align}

When all four indices coincide, the commutator involves various other generators. By applying the equation
 $S_{ii}=1-\sum_{k\ne i}S_{ik}$ implied by  the definition of the exchange
matrix \eqref{Sij},
one can obtain the following expressions from the most general commutators
\eqref{comFF}, \eqref{comFE}:
\begin{equation}
\label{comFiiFii}
[F_{ii},F_{ii}^+] =\, 4E_{ii}-\sum_{k\ne i}( E_{ii}+E_{kk}+E_{ik}+E_{ki})S_{ik}
+S_{ii}^2-(N-1)g^2,
\end{equation}
\begin{equation}
\label{comFiiEii}
[F_{ii},E_{ii}] =\, 2F_{ii}-\sum_{k\ne i}(F_{kk}+F_{ik})S_{ik}.
\end{equation}
In addition to the commutation relations, there exist various other algebraic relations among the generators.
For instance,  the products  $F_{ij}F_{kl}$ and
 $F_{ij}a_{k}$ are totally symmetric due to Eqs.~\eqref{com-apm}, \eqref{EF}. Similarly, the order of the last three indices
 as well as the order of the last three indices in  $E_{ij}F_{kl}$
  is not significant.
 Furthermore,  the expression  {\it $F_{ij}(E_{kl}+S_{kl})$ remains invariant} under cyclic
permutations of the same  indices  as the  $E_{ij}(E_{kl}+S_{kl})$ does [see Eq.~\eqref{EE-ES}].

Note that in the absence of the confining harmonic potential, the spectrum generating
condition \eqref{comHF} becomes trivial. According to Eq.~\eqref{Hw},
 one has to renormalise the related elements by $F_{ij}\to wF_{ij}$  in order to avoid
 singularities at the $w= 0$ limit. Hence, the spectrum generating
condition \eqref{comHF} turns to a trivial identity: $[\nabla^2,\nabla_i\nabla_j]=0$.

%

It is worth mentioning that the deformed symplectic algebra, generated by the elements \eqref{EF}  and the permutations,
can be considered as a subalgebra of the rational Cherednik algebra associated with
the symmetric group  \cite{etingof}.

\subsection{Conformal subalgebra}
The Dunkl-deformed symplectic algebra includes a conventional (nondeformed)
conformal subalgebra,
$sl(2,R)$, which comprises the following elements \cite{isakov}:
\begin{equation}
\label{Kpm3}
\begin{aligned}
&K_\pm=\frac12 \sum_iF^\pm_{ii}=\frac12 \sum_i a_i^{\pm2},
\\
&K_3=\frac12 H=\frac12\sum_iE_{ii}-\frac12S+\frac14N.
\end{aligned}
\end{equation}
They satisfy the standard commutation relation:
\begin{equation}
\label{comK}
[K_-,K_+]=2K_3,
\qquad
[K_3,K_\pm]=\pm K_\pm.
\end{equation}
These relations are derived from the algebra \eqref{com-apm}
and can also be obtained from Eqs.~\eqref{comFiiFkk}, \eqref{comFiiEkk},
\eqref{comFiiFii}, \eqref{comFiiEii}, and the relation
$$
\sum_i S_{ii}^2=N-2S.
$$

The conformal generators \eqref{Kpm3}  can be viewed as analogs
of those constructed long time ago
for the original Calogero model, without the inclusion of exchange operators
\cite{perel71}. The corresponding algebra differs also from the  conformal subalgebra
 of the  Virasoro algebra's  Dunkl deformation \cite{vasil}.
The operators  $K_\pm$ with excluded center of mass
have been considered in Ref.~\cite{jonke02}.

Instead of the lowering-raising generators, one can use
the elements $K_1$ and $K_2$ defined in a standard way:
\begin{equation}
\label{Kpm}
K_\pm=K_1\pm K_2.
\end{equation}
Then the defining relations \eqref{comK}
can be presented in a covariant form,
 highlighting the  equivalence with the  three-dimensional Lorentz algebra
$so(1,2)$:
\begin{equation}
\label{sl2}
[K_\alpha,K_\beta]=-\epsilon_{\alpha\beta\gamma}K^\gamma.
\end{equation}
Here, the $\epsilon$ represents the Levi-Civita tensor, and the index is raised by the Minkowski metric
$\gamma^{\alpha\beta}=\text{diag}(1,-1,-1)$.
The second element is anti-Hermitian,
while the remaining two are Hermitian:
$$
K_2^+=-K_2, \qquad
K_{1,3}^+=K_{1,3}.
$$
The first two generators are expressed via the Hamiltonian and
radial coordinate:
\begin{align}
\label{Kradial}
K_1
=\frac12(r^2-H),
\qquad
K_2=-\frac12r\partial_r -\frac{N}4.
\end{align}
Therefore, the conformal algebra commutes with the Dunkl angular
momentum:
$$
[K_\alpha, L_{ij}]=0.
$$

The replacement of $K_1$  by $\imath K_1$ in the equations
\eqref{sl2} reproduces the  familiar  relations of the $su(2)$ angular momentum
algebra in quantum mechanics.
The  Casimir element of the conformal algebra \eqref{sl2}
is  the square in  Minkowski space:
\begin{equation}
\label{cas}
K^2=
K^\alpha K_\alpha
=K_1^2-K_2^2-K_3^2.
\end{equation}
Up to a constant, it coincides with the angular
 Calogero Hamiltonian \eqref{Homega}, as can be easily verified:
\begin{equation}
\label{cas-ang}
H_\Omega=-4K^2-\left(\sfrac14 N-1\right)N.
\end{equation}

Note that when replacing $K_1\to -\imath K_1$, the conformal algebra commutations \eqref{sl2} turn
into the $su(2)$  relations, which describe the angular momentum or spin:
$[K_\alpha,K_\beta]=\imath \epsilon_{\alpha\beta\gamma}K_\gamma$ with
 Euclidian metric $\gamma^{\alpha\beta}\to -\delta^{\alpha\beta}$.
In particular,  the  $su(2)$ Lie algebra appears after introducing internal spin
degrees of freedom into the Calogero system \cite{minahan}.
Both algebras $so(1,2)$ and $su(2)$ are real  and
have the same complexification given by the Lie algebra $sl(2,C)$.
However, a mapping between them is not unitary,
and  they have essentially different properties in quantum physics.
In contrast to  the special unitary group, the conformal group is noncompact.
As a result, its  nontrivial unitary representations are infinite dimensional
while the $(2s+1)$-dimensional unitary $SU(2)$ multiplet combines the spin-$s$
states.

Finally note that the conformal \eqref{Kpm3} and Weyl \eqref{bpm}  generators form
together a closed algebra.
The nontrivial mixed commutation relations are given by
\begin{align}
[A^\mp,K^\pm]=\pm A^\pm,
\qquad
[K_3,A^\pm]=\pm \frac12A^\pm.
\end{align}

The Calogero model can be obtained by taking appropriate
 limits on the parameters of the Calogero-Sutherland type model
 (with or without confining potental) \cite{suther,inoz}.
On the other hand, the Hamiltonian and commuting integrals of the
Calogero-Sutherland model can be derived from those of the
Calogero-Moser system through a discrete reduction (folding) procedure
\cite{rev-poly}. However, this procedure is not safe
when applied to the conformal algebra.
Therefore, there is no straightforward extension of the spectrum generating algebra in general,
and the conformal algebra in particular, for trigonometric and hyperbolic potentials.

\section{Conformal group structure of Calogero wavefunctions}
\label{sec:wavefunctions}
This section is devoted to the presentation of conformal algebra on wave functions
of the (generalized) Calogero model.
In the first part, the known result on the deformed harmonic polynomials
are briefly described \cite{flp,DunklBook,Xu}.
The explicit form of the eigenfunctions  in spherical coordinates is given in the second subsection
both for the standard \cite{flp} and generalized Calogero models.
The main result  is obtained in the last part
where the lowest-weight infinite-dimensional $sl(2,R)$
multiplets are constructed  on the Hamiltonian's eigenfunctions.

\subsection{Deformed harmonic polynomials}

Recall that the unbounded generalized Calogero model (Calogero-Moser model with particle
exchanges)
may be expressed via the angular Hamiltonian
and  radial coordinate by inverting Eq.~\eqref{Homega'}:
\begin{equation}
\label{H0}
H_0=-\frac12 \nabla^2=-\frac12\partial_r^2  - \frac{N-1}{2r}\partial_r + \frac{H_\Omega}{2r^2}.
\end{equation}

Before proceeding further, let us briefly comment on the comparative properties of the three
particle-exchange models that have appeared so far \eqref{Hgen}, \eqref{Homega}, \eqref{H0}. The angular
Hamiltonian depends on the spherical coordinates and commutes  with the confined and unbound
generalized Calogero Hamiltonians: $[H_\Omega,H]=[H_\Omega,H_0]=0$. Therefore, the $H_\Omega$
has common eigenfunctions with the $H$ and $H_0$ separately, in which the angular and radial
coordinates are separated. The eigenfunctions of the confined system will be described
in the next subsection. The Hamiltonians $H$ and $H_\Omega$ describe bounded motion, and
each of them has a discrete spectrum.
In contrast, the   $H_0$ describes infinite motion with a continuous spectrum.
The fact that the transition from the $H$ to $H_0$ is not smooth was already
explained at the end of Sect.~\ref{sec:symmetry}.

Instead of Eq.~\eqref{H0}, we use
an equivalent Hamiltonian better suited
for the current section \cite{DunklBook}.
It  is obtained from the primary Hamiltonian through a similarity transformation
with respect to the $g$-th power of the Vandermonde polynomial's absolute value,
 \begin{equation}
 \label{phi}
 \phi=\phi(x)=\prod_{i<j}|x_i-x_j|^g,
 \end{equation}
participating in  the ground state's expression  \eqref{psi0}:
\begin{equation}
\label{H'0}
H'_0=\phi^{-1} H_0\phi
=-\frac12\nabla'\mbox{}^2
=-\frac12\partial^2
-g\sum_{i<j}\left[\frac{1}{x_i-x_j}(\partial_i-\partial_j) -\frac{1-s_{ij}}{(x_i-x_j)^2}\right].
\end{equation}
Under such a transformation, the Dunkl operator, in turn, undergoes the following shift:
\begin{equation}
\label{du'}
\nabla'_i=\phi^{-1}\nabla_i\phi=\partial_i+\sum_{j\ne i}\frac{g}{x_i-x_j}(1-s_{ij}).
\end{equation}
Evidently, the applied similarity map is not unitary. As a consequence, both $H'_0$
and $\pi'_i=-\imath \nabla'_i$ are not Hermitian.

An advantage of the shifted Dunkl operator over the initial
one \eqref{du} is the reduction to the  conventional    derivative
when applied to a  symmetric function.
Futhermore,  the  deformed Laplace  equation
\begin{equation}
\label{harm}
\nabla'^2h(x)=0
\end{equation}
  has polynomial solutions, which  are  Dunkl deformations of the usual harmonic polynomials
\cite{dunkl,DunklBook}.
The wavefunctions of the angular momentum squared are expressed
using such polynomials.
In fact, it is easy to see that
the relation \eqref{Homega'}  implies  that any deformed
homogeneous harmonic polynomial is an
eigenfunction  of the shifted angular Calogero Hamiltonian.

Consider a subset among the deformed homogeneous harmonic polynomials given by the following concise
expression \cite{Xu}:
\begin{equation}
\label{hn}
h_{\bf n}(x)=r^{2(E_0+m_{\bf n}-1)}
\nabla'^{n_1}_1\nabla'^{n_2}_2\dots \nabla'^{n_N}_N
r^{2(1-E_0)}
\end{equation}
where $E_0$ was already defined \eqref{E0}, and  the polynomial's degree takes on
the value
\begin{equation}
\label{mn}
m_{\bf n}=\sum_{i=1}^N n_i,
\qquad
r\partial_r h_{\bf n}(x)=m_{\bf n} h_{\bf n}(x).
\end{equation}
The functions $h_{\bf n}(x)$ form a basis in the space of the
 deformed homogeneous harmonic polynomials of degree $m_{\bf n}$. However, this basis, in general,
 is overcomplete due to the linear dependence among the corresponding elements [see Eq~\eqref{dep} below].

The Dunkl derivative obeys a modified Leibniz rule:
\begin{equation}
\label{leibniz}
\nabla'_i(\phi\psi)=\phi\nabla'_i \psi + \psi\nabla'_i \phi
- g\sum_{k\ne i}\frac{(\phi-s_{ik}\phi)(\psi-s_{ik}\psi)}{x_i-x_k},
\end{equation}
where $s_{ij}\phi(x)=\phi(s_{ij}x)$. The sophisticated  third term
in Eq.~\eqref{leibniz} vanishes if one of the two functions is symmetric, say $\phi$: $s_{ij}\phi=\phi$,
giving rise to the standard Leibniz rule. Then,  the condition $\nabla'^2 h_{\bf n}(x)=0$ is
verified by straightforward calculations.

The symmetrization procedure over all coordinates reduces the deformed harmonic  polynomials
\eqref{hn} to the expression \cite{flp}
\begin{align}
\label{hk}
h_{\bf k}^\text{sym}(x)&=r^{2(E_0+m_{\bf k}-1)}
\mathcal{D}_1^{k_1}\mathcal{D}_3^{k_3}\dots \mathcal{D}_N^{k_N}
r^{2(1-E_0)}.
\end{align}
Here and below we use the implicit convention that ${\bf k}$ (with $k_2=0$)
denotes the modes of symmetric polynomials, and $\bf n$ denotes those
of asymmetric polynomials \eqref{hn}.
We have used also a shortened notations for the Newton power sums in the shifted Dunkl operators:
$$
\mathcal{D}_l=\sum_{i=1}^N \nabla'_i\mbox{}^l.
$$
Symmetrization projects out the operators to their permutation-invariant counterparts.
As a result, the entire algebra restricts itself to a part referred to in the mathematical literature
as a spherical subalgebra of the rational Cherednik algebra \cite{etingof}.
Note that the operators in Eqs.~\eqref{hn}, \eqref{hk} act on the deformed harmonic function:
\begin{equation}
\label{r-a}
\mathcal{D}_2 r^{-2(E_0-1)}=\nabla'^2 r^{-2(E_0-1)}=0.
\end{equation}
Therefore, the Dunkl Laplacian $\mathcal{D}_2$ is absent in the expression of the polynomial \eqref{hk}.
Its degree is  defined by Eq.~\eqref{Ekb}:
\begin{equation}
\label{ord-hk}
r\partial_r h_{\bf k}^\text{sym}(x)=m_{\bf k} h_{\bf k}^\text{sym}(x)
\qquad
\text{with}
\qquad
k_2=0.
\end{equation}

Note that, due to the condition \eqref{ord-hk},  the deformed harmonic
polynomials \eqref{hn} are not linearly independent \cite{Xu}:
\begin{equation}
\label{dep}
\sum_{i=1}^N h_{{\bf n}+2{\bf e}_i}(x)=0
\end{equation}
with the standard notation for the unit vectors \eqref{A1}.

\subsection{Wavefunctions in deformed harmonic polynomials}
First, let us consider the Dunkl angular momentum squared, which coincides with  the
angular part of the generalized Calogero model \eqref{Homega}.
The related eigenfunctions  are  the product of the deformed harmonic polynomial \eqref{hn}
and shift function \eqref{phi} as explained above.
The energy levels  are obtained immediately using the expression for the
angular Hamiltonian \eqref{Homega'}:
\begin{align}
\label{phihn}
&H_\Omega\, \phi(x) h_{\bf n}(x)=\mathcal{E}_\beta\phi(x) h_{\bf n}(x),
\\
\label{epsilon}
&{\cal E}_\beta=\beta^2-(\sfrac12 N-1)^2.
\end{align}
They depend on a single quantum number expressed through  the degree  of the polynomial \eqref{mn}:
\begin{equation}
	\label{beta}
\beta=\beta_m=E_0+m-1,
\qquad
m=m_\textbf{n}=\sum_{l=1}^N n_l,
\end{equation}
where $E_0$ is the lowest energy of the
Calogero model \eqref{Ekb}.
This property leads to the highest
 level of degeneracy and maximal superintegrability of the angular system \cite{flp,CalCoul}.
Note that the eigenstates are not independent but subjected to a constraint
inherited from the condition \eqref{dep}.

Since we are dealing with the angular system, the wavefunctions  have to depend
only on the angular coordinates $u_i=\cos\theta_i=x_i/r$ located on unit  $d$-sphere with $d=N-1$: $\sum_i u_i^2=1$.
Indeed, the radial coordinate may be eliminated after canceling out the common factor  $r^{\beta-\frac12 N +1}$
from both sides of Eq.~\eqref{phihn}.

 Using the wavefunctions
of the angular Hamiltonian,
the energy eigenfunctions of  the (generalized) Calogero model
may be constructed in spherical coordinates. This fact reflects a common feature of superintegrable
Hamiltonians, which allows separation of variables   in a few coordinate systems.
The simplest and most conventional solution of the generalized Calogero model is given
in terms of the Cartesian coordinates \eqref{En}.

First, note that the standard similarity transformation with respect to the function $r^{\frac12(N-1)}$,
when applied to the generalized Calogero (Calogero-Moser)
system \eqref{H0},  absorbs the term that is linear in the radial derivative.  As a result, the
Hamiltonian acquires the following simple form:
\begin{equation}
\label{Hshift}
\begin{aligned}
	r^{\frac12(N-1)}Hr^{-\frac12(N-1)}&=-\frac12\partial_r^2   + \frac12 r^2
	+ \frac{H_\Omega+\varepsilon}{2r^2},
\\
\varepsilon&=\sfrac14(N-1)(N-3).
\end{aligned}
\end{equation}
The deformed spherical harmonics \eqref{phihn}, \eqref{epsilon} are eigenfunctions
of the shifted Dunkl angular momentum squared operator, $H_\Omega+\varepsilon$,
with the eigenvalue
$$
{\cal E}_\beta+\varepsilon=\beta^2-\sfrac14.
$$
So, the angular and radial coordinates of the transformed
  Hamiltonian \eqref{Hshift} are separated.
The radial part describes a one-dimensional singular oscillator
 with the spectrum and eigenfunctions, respectively, given by
\begin{equation}
\label{Ek}
 E_k=\beta+2k+1,
 \qquad
e^{-\frac{r^2}{2}}r^{\beta+\frac12}L_k^\beta(r^2)
\end{equation}
with $k=0,1,2,\dots$, and
$$
L_k^{\beta}(z)=\frac1{k!}z^{-\beta}e^z\frac{d^k}{dz^k}\left(e^{-z}z^{k+\beta}\right)
$$
being the $k$-th order associated  Laguerre polynomial   \cite{rev-olsh}.



Hence, the eigenstates of the generalized Calogero model in terms of the spherical harmonics
are represented by the following functions:
\begin{align}
\label{psik}
\Psi_{{\bf n},k}(x)&=e^{-\frac{r^2}{2}}r^{\beta-\frac12N+1}L_k^\beta(r^2)\phi(u) h_{\bf{n}}(u)
\\
\label{psik'}
&=e^{-\frac{r^2}{2}}L_k^\beta(r^2)\phi(x) h_{\bf{n}}(x),
\end{align}
where $u_i$ are  the  angular coordinates defined above.
Here, the radial quantum number $k$ counts the nodes of the radial part of the wavefunction
corresponding to the zeros of the polynomial
$L_k^\beta(z)$ in the positive region.
The corresponding
energies are obtained from Eqs.~\eqref{beta} and \eqref{Ek}:
\begin{equation}
\label{Enk}
E_{{\bf n},k}
=E_0+m+2k=
E_0+\sum_{l=1}^{N} n_l+2k.
\end{equation}
%

By selecting the symmetric  harmonic polynomials \eqref{hk} instead of the standard ones,
one obtains states for indistinguishable bosons that are also  eigenfunctions
of the conventional Calogero model \cite{flp},
\begin{equation}
\label{psisym}
\Psi^\text{sym}_{k_1k_2k_3\dots k_N}(x)=e^{-\frac{r^2}{2}}L_{k_2}^\alpha(r^2)\phi(x) h^\text{sym}_{k_10\,k_3\dots k_N}(x).
\end{equation}
Now,  the radial quantum number, which corresponds to the lower index of the associated Laguerre polynomial,
coincides  with the second quantum number that describes the state.
The parameter $\alpha$ is similar to  $\beta$ but is associated with
the degree of the symmetric homogeneous polynomial \eqref{hk}, \eqref{ord-hk}:
\begin{equation}
\label{alpha}
	\alpha=\alpha_m=E_0+m-1,
	\qquad
m= m_{\bf k} =k_1+\sum_{l=3}^N l k_l.
\end{equation}

The energy spectrum is determined by the
same formula as in the case of Cartesian coordinates \eqref{Ekb}:
\begin{equation}
\label{Ek-alpha}
E_{\bf k}=E_0+\sum_{l=1}^N l k_l=\alpha+2k_2+1.
\end{equation}
However, in general,  the quantum numbers $k_l$ that characterize the energy
eigenstates in the Cartesian
\eqref{ksym} and spherical \eqref{psisym} coordinate systems differ mutually.
This discrepancy arises from the degeneracy of energy levels,
reflecting the superintegrability of the Calogero model.

In comparison with the spectrum of the generalized Calogero
model \eqref{Enk}, the second quantum number in the spectrum
\eqref{Ek-alpha}
 is exclusively determined by the radial part of the
wavefunction, given by the associated Laguerre polynomial
\eqref{psisym}.
Later, it will be shown that this quantum number is controlled by the conformal group.

Finally, it is worth mentioning that a recent detailed study of wavefunctions
expressed in terms of deformed spherical harmonics has been conducted
for the generalized Calogero model based on the reflection groups $B_2$ and $H_3$,
which correspond to the symmetries of the square and icosahedron, respectively \cite{dunkl22-23}.


\subsection{Conformal group action on wavefunctions in spherical coordinates}
\label{sec:radial}

The symmetric wavefunctions of the Calogero model in spherical coordinates \eqref{psisym} with fixed values
of all quantum numbers $k_l$ except the
second one (with $l=2$) are organized into the infinite-dimensional lowest-weight representation of the conformal
algebra.
When applying the $SL(2,R)$ generators to these wavefunctions,
the following  expressions are obtained:
\begin{align}
\label{K-psi}
K_-\Psi^\text{sym}_{\dots k_2\dots  } &=-(\alpha+k_2)\Psi^\text{sym}_{\dots k_2-1\dots},
\\
\label{K+psi}
K_+\Psi^\text{sym}_{\dots k_2-1\dots } &=-k_2\Psi^\text{sym}_{\dots k_2\dots },
\\
\label{K3psi}
K_3\Psi^\text{sym}_{\dots k_2\dots } &
=\sfrac12 E_{\bf k}\Psi^\text{sym}_{\dots k_2\dots },
\end{align}
where  the condition $\Psi^\text{sym}_{\dots k_2\dots  }=0$ is imposed for the negative values of $k_2$. Here, the dots represent
 the remaining
quantum numbers ($k_l$ with $l\ne 2$) that are not influenced by the conformal group.
According to the definition of the conformal algebra \eqref{Kpm3},
the eigenvalue of the diagonal element  equals  half the state's energy \eqref{Ek-alpha}.
Since the normalization constants in front of the wavefunctions are neglected,
the matrices of the operators $K_\pm$ and $K_3$ are not in their conventional form
used in quantum mechanics.

%
The relations \eqref{K+psi} and \eqref{K-psi} follow from
the radial representation of the conformal generators
\eqref{Kpm}, \eqref{Kradial}, the spectrum expression \eqref{Ek-alpha},  as well as
the following recurrence relation among the associated Laguerre
polynomials:
\begin{align*}
(z\partial_z -k)L_k^\alpha (z)&=-(\alpha + k)L_{k-1}^\alpha (z),
\\
(z\partial_z - z + \alpha + k)L_{k-1}^\alpha (z)&=kL_k^\alpha (z)
\end{align*}
which holds for any  integer $k\ge 1$.

The infinite-dimensional $SL(2,R)$ multiplet given by the action \eqref{K-psi}--\eqref{K3psi} is generated by
a wavefunction with  $k_2=0$,
$$
\Psi^\text{sym}_{k_1 0 k_3 \dots  }(x)=e^{-\frac{r^2}{2}}\phi(x) h^\text{sym}_{k_10\,k_3\dots k_N}(x),
$$
which satisfies the  lowest-state condition
$$
K_-\Psi^\text{sym}_{k_1 0 k_3 \dots  }=0 ,
\qquad
K_3\Psi^\text{sym}_{k_1 0 k_3 \dots  }=s\Psi_{k_1 0 k_3 \dots  }.
$$
Here, the conformal spin  depends on all quantum numbers, see Eq.~\eqref{alpha}: 
\begin{equation}
\label{s}
	s=\sfrac12(\alpha+1).
\end{equation}

The Casimir element of the conformal algebra \eqref{cas} assumes a constant value in
 the present representation:
\begin{equation}
\label{Ksq}
K^2=-\sfrac14(\alpha^2-1)=-s(s-1).
\end{equation}
The relations \eqref{cas-ang}  between the Casimir elements of the conformal algebra and Dunkl
angular momentum  being applied to the discussed wavefunctions leads
to an explicit relation between the corresponding constants, as  given by Eq.~\eqref{epsilon}
with $\beta$ replaced by $\alpha$.

%

In summary, only the radial (second) quantum number of the symmetric wavefintion
 \eqref{psisym} 
is influenced by the conformal group in the Calogero model. This property is due to
the choice of spherical coordinates and remains true also for asymmetric wavefunctions
which will be considered below in this context.
In contrast, in Cartesian coordinates, conformal transformations can also affect other quantum numbers.
To be more specific, the Cartesian basis formed by the symmetric wavefunctions  \eqref{ksym}
exhibits similar behavior to the spherical basis  \eqref{psisym}
when subjected to the raising and diagonal generators:
%
%
\begin{align*}
	K_+|{\bf k}\rangle_\text{sym}=\sfrac12|{\bf k}+{\bf e}_2\rangle_\text{sym},
\qquad
	K_3|{\bf k}\rangle_\text{sym}=\sfrac12 E_{\bf k}|{\bf k}\rangle_\text{sym}.
\end{align*}
However, the lowering generator alters all quantum numbers. For instance, if
 $k_l=0$ for all $l\ge 3$, one gets:
\begin{align*}
K_-|k_1,k_2\rangle_\text{sym}=2k_2(\alpha+ k_2)|k_1,k_2-1\rangle_\text{sym}
+(k_1-1)|k_1-2,k_2\rangle_\text{sym},
\end{align*}
where the vanishing quantum numbers are excluded within the ket states.
If the last term is omitted, the representation above becomes equivalent to the previous one \eqref{K-psi}--\eqref{K3psi}
 through a similarity map:
$$
|k_1,k_2\rangle_\text{sym}\to (-1)^{k_2}k_2!\Psi^\text{sym}_{\dots k_2\dots }.
$$
Of course, when using Jacobi coordinates, it is possible to separate
the center of mass and  associated quantum number, $k_1$.
 However, the higher-level quantum numbers, starting from $k_3$,
  will remain coupled. 

Similar types of $SL(2,R)$ representations appear in the generalized Calogero model with particle exchanges
\eqref{Hgen}, which has more general, nonsymmetric eigenstates
$\Psi_{{\bf n},k}$ \eqref{psik'}.
The corresponding multiplets are
generated from the ground state
$
\Psi_{{\bf n},0}(x)=e^{-\frac{r^2}{2}}\phi(x) h_{\bf n}(x)
$
and are characterized by the conformal spin given by the formula \eqref{s}
with the parameter $\alpha$ substituted by $\beta$ \eqref{beta}:
\begin{equation}
\label{sbeta}
s=\sfrac12(\beta+1).
\end{equation}
The eigenvalue associated with
the diagonal generator is equal to half of the energy
which is now determined by Eq.~\eqref{Enk}:
\begin{equation}
\begin{aligned}
	K_-\Psi_{{\bf n},k } &=-(\beta+k)\Psi_{{\bf n},k -1},
	\qquad
	K_+\Psi_{{\bf n},k-1 } &=-k\Psi_{{\bf n},k },
	\\
	K_3\Psi_{{\bf n},k } &=\sfrac12E_{{\bf n},k}\Psi_{{\bf n},k }.
\end{aligned}
\end{equation}
It depends on the radial quantum number $k$ which also distinguishes
individual states within  a single multiplet.
The quantum numbers $n_i$, which characterize the spherical
harmonics, remain unaffected by the conformal transformation.
They determine  the value of the conformal spin
\eqref{s'}, as shown in Eq.~\eqref{beta}.

\section{Conformal group structure of Calogero-Moser integrals}
\label{sec:integrals}

\subsection{Rotated $sl(2,R)$ basis}
In the previous section, we discussed the behavior of the spectrum and wavefunctions
of the Calogero model, both with and without particle exchanges,
under the action of the conformal group.
The latter is a part of a more general dynamical symmetry formed by the Dunkl symplectic group.
In the current section, we study the conformal multiplets generated by the quantum analogs of
Liouville integrals
 by adapting the method used in Ref.~\cite{hak14} to the exchange-operator approach.
The additional constants of motion are then obtained through the tensor-product decomposition,
which is equivalent to the usual angular momentum sum rule in quantum mechanics.

Since both  the second and third generators of the conformal algebra
have a space-like signature \eqref{cas}, a rotation in this plane
results in an equivalent algebra \eqref{sl2}. In particular, the $\pi/2$
rotation along the first conformal component maps the
initial generators  to their modified counterparts:
\begin{equation}
\label{K'}
K'_\alpha=e^{\frac\pi2 K_1}K_\alpha e^{-\frac\pi2 K_1}.
\end{equation}
Clearly, the similarity transformation above does not affect the generator $K_1$
but it exchanges the two other components as follows:
\begin{equation}
\label{K'23}
 K'_2=-K_3,
\qquad
 K_3'=K_2.
\end{equation}
As a result, the rotated lowering and raising
$SL(2,R)$ generators \eqref{Kpm} acquire
the following explicit form:
\begin{equation}
\label{K'pm}
\begin{aligned}
K'_+=K_1-K_3=\sfrac12 \nabla^2=-H_0,
\qquad
K'_-=K_1+ K_3=\sfrac12 r^2.
\end{aligned}
\end{equation}
The conformal algebra representation via the Dunkl operators
first appeared in above form in Ref.~\cite{heckman}.
Recall that the observable
$H_0=-\sfrac12I_2$ represents the generalized Calogero-Moser Hamiltonian \eqref{H0}.

Note that the  transformation \eqref{K'} is not unitary, so
both $K'_\pm$ are Hermitian, while $K_\pm$ are conjugates of each other.
In fact, the same rotation  relates the
generators of the deformed Weyl and Dunkl algebras,
$$
a_i\to \nabla_i,
\qquad
 a_i^+\to x_i.
$$
As a result, the generalized Calogero Hamiltonian \eqref{Ha} transforms
to
$$
H\to -2K'_3 =r\partial_r+\frac12N,
$$
while the ground state maps to   $\phi$ \eqref{phi},
which
vanishes under the action of the Dunkl operator: $\nabla_i \phi(x)=0$.

\subsection{Conformal multiplets generated by quantum counterparts of classical Liouville integrals}
The commuting invariants of the unbounded Calogero (Calogero-Moser)  model,
which correspond to the classical integrals, are
symmetric polynomials in Dunkl operators, typically represented as power sums
 \cite{poly92}:
\begin{equation}
	\label{In}
	I_n=\sum_{i=1}^N \nabla_i^n.
\end{equation}
In fact, the valid quantities must be Hermitian and expressed via
the Dunkl momentum, $\sum_i \pi _i^n=(-\imath)^n I_n$. Here,
 we disregard this requirement to avoid imaginary units in the
 formulas below.

The adjoint action of the conformal algebra
on the above integral has a
quite simple form in the newly defined basis. Indeed, the
third generator is diagonal while
the raising  one just kills it:
\begin{equation}
\hat K'_+I_n=0,
\qquad
\hat K'_3I_n=\sfrac12 nI_n,
\end{equation}
where the hat means the commutator:
$\hat Xf=[X,f]$.

It is easy to see that the integral \eqref{In} generates the highest-weight irreducible
 representation of the $SL(2,R)$ group with a conformal spin
\begin{equation}
\label{s'}
s'=\sfrac12 n.
\end{equation}
The multiplet is composed of states $I_{n,l}$
with  $l=0,1,\dots, n$, and the state $I_{n,0}=I_n$ corresponds to the highest-weight one:
\begin{equation}
\label{Inl}
I_{n,l}= (\hat K'_-)^lI_n,
\qquad
\hat K'_3I_{n,l}=(s'- l)I_{n,l}.
\end{equation}
The raising generator acts on these states
in a standard way:
\begin{equation}
\label{HInl}
\hat K'_+I_{n,l}=-\hat H_0 I_{n,l}=-l(n-l+1)I_{n,l-1}.
\end{equation}

In particular, the first integral $I_1=\sum_i\partial_i$, which is proportional to
the total momentum, generates
a doublet ($s'=1/2$) with the lowest function proportional to the center-of-mass coordinate
$I_{1,1}=-\sum_ix_i$.
The second integral generates a triplet ($s'=1$) consisting of the following members:
\begin{equation}
\label{triplet}
\begin{gathered}
 I_2=2K'_+,
 \qquad
 I_{2,1}=4K'_3=-2r\partial_r- N,
\qquad
I_{2,2}=4K'_-=2r^2.
\end{gathered}
\end{equation}

The square of the conformal spin \eqref{cas} assumes
a constant value within a single multiplet:
\begin{equation}
\label{K'sq}
\hat K^2=-\sfrac14n(n+1)=-s'(s'+1).
\end{equation}
%
Note that the values \eqref{Ksq} and \eqref{K'sq} are mapped to each other
under the substitution $s'\to-s$.

\subsection{Descendants of commuting integral in Weyl-ordered form}

The conformal-algebra descendants of the commuting integral of motion
\eqref{In} are expressed via
the Weyl-ordered operator product
of the coordinate and Dunkl operator
in the following way:
\begin{equation}
\label{IW}
I_{n,l}=(-1)^l \frac{n!}{(n-l)!}\sum_{i=1}^N\left(x_i^l\nabla_i^{n-l}\right)_W.
\end{equation}
Here, the Weyl ordering  is used, which symmetrizes over all  possible orderings in the operator product
and appears usually in quantum field theory.
For the product of powers of two operators $a$ and $b$,  it is given by the
expression:
\begin{equation}
\label{albk}
(a^lb^k)_W=\sbinom{k+l}{l}^{\scriptscriptstyle -1}(a^lb^k+a^{l-1}bab^{k-1}+\dots+b^ka^l).
\end{equation}
Another definition  is provided by the generating function:
\begin{equation}
\label{weyl}
( a+ v b)^n=\sum_{l=0}^n\binom{n}{l}v^l \left(a^{n-l}b^{l}\right)_W.
\end{equation}
Notice that  the symmetrized  product \eqref{albk}  appeared
in the  Lax and Dunkl ladder operators
\cite{wadati,naray}.


Eq.~\eqref{IW} follows immediately from the commutation relations:
$$
\hat K'_-\nabla_i=-x_i, \qquad
\hat K'_- x_i=0.
$$
Note that the factorials  in Eq.~\eqref{IW} contract with the binomial factor in front
of the Weyl product \eqref{albk}, resulting in an overall $l!$ factor.

In particular, the first descendant from the family  has the following explicit form:
\begin{align}
\label{I1}
I_{n,1}
&=-\sum_{i=1}^N\sum_{k=0}^{n-1}\nabla_i^kx_i\nabla_i^{n-k-1}.
\end{align}
At the same time, the last one reduces
to the center of mass:
 $$
 I_{n,n}=(-1)^n n!\sum_i x_i.
 $$

Note that a generating polynomial  encompassing all the descendants
\eqref{IW} can be constructed as a Newton's power sum in an appropriate
superposition of the Dunkl operator with
corresponding coordinate  \eqref{weyl}:
$$
I_n(v):=\sum_{l=0}^{n} (-v)^lI_{n,l}=\sum_{i=1}^N (\nabla_i-v x_i)^n.
$$

It should also be noted that soon after the first version of the current article was sent to arXiv,
an interesting investigation appeared \cite{correa24}. This investigation utilized the Weyl-ordered
product for the construction of permutation-invariant observables of the generalized three-particle Calogero model,
the study of their algebra, and the derivation of related Casimir invariants.

\subsection{Additional integrals from product multiplets}

As mentioned in Introduction, the Calogero-Moser model is superintegrable.
In addition to the commuting integrals $I_n$, it possesses additional constants of motion.
Together, these constitute a set of $2N-1$ independent quantities
that commute with the Hamiltonian \cite{woj83,kuznetsov,gonera98}.
Actually, superintegrability  is closely
 related to  the dynamical $SL(2,R)$ symmetry, which
 can be used to derive the additional integrals from the commuting ones \cite{gonera98,hak10,hak14}.

Alternatively, the additional integrals can be recovered as the highest-weight functions in the product
representation \cite{hak14}.
Indeed, a symmetric bilinear combination of two invariants from the
commuting family \eqref{In},
\begin{equation}
\label{Isym}
\{I_{n_1},I_{n_2}\}=I_{n_1}I_{n_2}+I_{n_2}I_{n_1},
\end{equation}
possesses a structure of the tensor product representation with respect to the
  $sl(2,R)$ algebra's adjoint action.
The symmetrized invariant \eqref{Isym} decomposes according to the usual momentum sum rule
in quantum mechanics:
$(s_1)\otimes (s_2)=(s_1+s_2)\oplus (s_1+s_2-1)\oplus\dots\oplus (|s_1-s_2|)
$.
Apparently, the highest-weight states  are
 integrals of motion:
\begin{align*}
I^{n_1,n_2}_{n_1+n_2-2k}=&\sum_{l=0}^k
\frac{(-1)^l}{2} \binom{n_1-k+l}{l} \binom{n_2-l}{k-l}\{I_{n_1,k-l}, I_{n_2,l}\},
\\
\hat K'_+I^{n_1,n_2}_{n_1+n_2-2k}=&\big[H_0,I^{n_1,n_2}_{n_1+n_2-2k}\big]=0
\end{align*}
where  $0\le k \le |n_1-n_2|$. This integral is parameterized by  twice the spin value
of the conformal multiplet it generates:
$$
\hat K'_3I^{2s_1,2s_2}_{2s}=sI^{2s_1,2s_2}_{2s}
\quad\text{with}\quad
s=s_1+s_2-k.
$$
For simplicity, we omit the prime symbol
in the  spin's notation \eqref{s'}.

The first nontrivial integral
within this  family corresponds to the $k=1$
case:
\begin{equation}
\label{In1n2}
I^{n_1,n_2}_{n_1+n_2-2}=
s_2\{I_{n_1,1}, I_{n_2}\}-s_1\{I_{n_1}, I_{n_2,1}\}.
\end{equation}

In case  when the  second multiplet is a triplet, defined  in Eq.~\eqref{triplet}, the integral
of motion can also be expressed as a commutator with the conformal spin's square,
or the generalized angular Calogero Hamiltonian:
\begin{align*}
I^{n,2}_{n}=\,\{I_{n,1}, K'_+\}-2n\{I_{n},K'_3\}
=\,2[K^2,I_n]=-\frac12[H_\Omega,I_n].
\end{align*}
These equations are derived easily from the expression
$K^2=\frac12\{K'_+,K'_-\}-K'^2_3$ for the conformal spin's square and the relation
\eqref{cas-ang}.


\subsection{Generalized Calogero-Moser's integrals}

The construction of additional integrals of motion based on the dynamical $SL(2,R)$
symmetry is even more transparent for the Calogero-Moser model with particle exchanges.
In this case, the commuting  integrals are formed by the Dunkl momentum components
$\pi_i=-\imath \nabla_i$ without any symmetrization
procedure. 
Each component of the Dunkl operator generates a conformal doublet ($s'=1/2$) with
\begin{equation}
\label{I1gen}
I_1=\nabla_i  \qquad \text{and} \qquad I_{1,1}=-x_i.
\end{equation}
According to the momentum sum rule,  a pair of such multiplets,
generated by  $I_1=\nabla_i$ and  $I'_1=\nabla_j$,
produce a singlet ($s'=0$).
The latter corresponds to a conserved quantity, which coincides with  the Dunkl angular
momentum component \eqref{Lij}:
\begin{equation}
\label{I11gen}
I^{1,1}_{0}=\frac12\{I_{1,1},I'_1\} - \frac12\{I_1,I'_{1,1}\}
=\imath L_{ij}.
\end{equation}
Therefore, the Dunkl angular momentum symmetry is inherited from the
Dunkl momentum and dynamical conformal symmetries.

The symmetric polynomials in integrals of the generalized Hamiltonian
\eqref{H0} are constants of motion of the standard Calogero-Moser model.
In particular, after symmetrization procedure, two integrals $\nabla_i$ and $\nabla_j$
reduce to the same expression $I_1$ from \eqref{In} while the two others, $I_{1,1}$, $I'_{1,1}$
from Eqs.~\eqref{I1gen},
become proportional to the center of mass.
Therefore, the integral $I^{1,1}_{0}$ will reduce to zero, which is
clear also due to antisymmetric nature of the angular momentum tensor \eqref{I11gen}.
Taking the powers  $\nabla^{n_1}_i$, $\nabla^{n_2}_j$
instead, one arrives at the integrals \eqref{In1n2}.

 \section{Conclusion}
 It is well known that
the Weyl algebra $w(N)$ generated by  the creation-annihilation operators
and  the symplectic algebra $sp(2N)$ formed  by their  bilinear combinations
generate a largest possible finite-dimensional group which generates the
spectrum of the $N$-dimensional isotropic oscillator.
In the current article,  the spectrum generating algebra was deformed in case of
the presence of additional inverse square (Calogero) potential.
The construction was based on the exchange operator formalism, in which the
derivatives in observables are replaced by the Dunkl operators.
%
 In particular, the commutation relations among the deformed  generators
are derived explicitly and presented in compact form.

The Dunkl analog of the symplectic algebra  contains  the
deformed unitary subalgebra which combines the symmetries of the generalized Calogero
Hamiltonian \cite{fh}. The remaining, spectrum generating part maps in-between the
different energy levels and involves  the standard
$sl(2,R)$ conformal subalgebra. A simple relation is obtained between the Casimir elements
of both the conformal spin and
Dunkl angular momentum, given by (modified) squares in the corresponding generators.
This correspondence has suggested to  analyze the conformal structure of the Calogero
wavefunctions   in the spherical coordinates based
on the Dunkl analog of the spherical harmonics. It turned out that, in fact,
the second quantum number   is driven by the conformal group, and
each eigenstate, in which it vanishes,  generates the infinite dimensional lowest-weight
 $sl(2,R)$ multiplet.

Next, a $\pi/2$ rotation has been applied to the
 conformal generators in order to demonstrate that the $n$-th
 integral of motion from the commuting family of the (generalized) Calogero-Moser model
 generates the finite-dimensional (non-unitary)  highest weight spin-$\frac n2$ $sl(2,R)$ multiplet.
The descendants of this representation  have been
expressed in terms of  the Weyl-ordered product in quantum field theory.
The highest states  in a product of two or more such multiplets
produce additional integrals of motion.

\acknowledgments
The author is grateful to M.~Feigin for preliminary discussions.
 The work was supported by the Armenian Science Committee Grants
Nos. 21AG-1C047 and 24FP-1F039.

\end{document}